\documentclass[tightenlines,twocolumn,showpacs,superscriptaddress,prb]{revtex4}
\usepackage{graphicx}
\usepackage{epsfig}

\begin{document}

\title{Semimetallic Antiferromagnetism in the Half-Heusler Structure: CuMnSb} 

\author{T. Jeong}
\affiliation{Department of Physics, University of California, Davis CA 95616}
\author{Ruben Weht}
\affiliation{Departamento de F\'{\i}sica, CNEA,
Avda. General Paz y Constituyentes, 1650 - San Mart\'{\i}n, Argentina}
\author{W. E. Pickett}
\affiliation{Department of Physics, University of California, Davis CA 95616}

\date{\today}

\begin{abstract}
The half-Heusler compound CuMnSb, the first
antiferromagnet (AFM) in the Mn-based class of Heuslers and half-Heuslers
that contains several conventional and half metallic ferromagnets, shows
a peculiar stability of its magnetic order in high magnetic fields. 
Density functional based studies reveal an unusual nature of its
unstable (and therefore unseen) paramagnetic state, which for one electron
less (CuMnSn, for example) would be 
a zero gap semiconductor (accidentally so) between two sets of very narrow,
topologically separate bands of Mn $3d$ character.  The extremely flat
Mn $3d$ bands result from the environment: Mn has four 
tetrahedrally coordinated Cu atoms whose $3d$ states lie well below the Fermi 
level, and the other four tetrahedrally coordinated sites are empty, 
leaving chemically isolated Mn $3d$ states.   
The AFM phase can be pictured heuristically
as a self-doped Cu$^{1+}$Mn$^{2+}$Sb$^{3-}$
compensated semimetal with heavy mass electrons and light mass holes,
with magnetic coupling proceeding through Kondo and/or antiKondo
coupling separately through the two carrier types.
The ratio of the linear specific heat
coefficient and the calculated Fermi level density of states indicates
a large mass enhancement $m^*/m \sim 5$, or larger if a correlated
band structure is taken as the reference.
\end{abstract}


\maketitle

\section{Introduction}

The study of unusual magnetic materials has increased dramatically
in the past decade, due to new phenomena such as colossal magnetoresistance,
half-metallic ferromagnetism, spin gaps and novel heavy fermion
behavior, to name a few.  The Heusler (AT$_2$X type) and 
half Heusler (ATX type, T = transition metal, X = metal or metalloid) 
structures have
produced a number of unusually interesting compounds.  For example,
ferromagnetic (FM) PtMnSb was discovered\cite{engen} to display the largest
known magneto-optic Kerr effect at room temperature (at
photon energy  $\hbar \omega$ = 1.75 eV). A number of them have been
proposed as half metallic ferromagnets, and NiMnSb (with one more electron
than CuMnSb) is one of the most
widely accepted examples of half metallic ferromagnetism.

We address here
the half-Heusler compound CuMnSb, which has its own peculiarities.  
In a class of
intermetallic magnets where ferromagnetism (FM) can occur as high at
1000 K, CuMnSb
develops {\it anti}ferromagnetism (AFM)
at the relatively low temperature, with reports of
T$_N$~=~55~K from Endo,\cite{endo} 62 K from 
Helmholdt {\it et al.},\cite{helmholdt} and 50 K from Boeuf.\cite{thesis}
A large value of the ordered Mn moment
$m$ = 3.9-4.0 $\mu_B$ has been obtained from neutron scattering 
measurements.\cite{helmholdt,forster}
This moment suggests that a ``monovalent Mn'' may provide
a good picture, with $d^5_{\uparrow}, d^1_{\downarrow}$ 
high spin configuration.
The Curie-Weiss moment is large, reported variously as 5.4
$\mu_B$~[\onlinecite{endo}], 6.3 $\mu_B$[\onlinecite{thesis}]
and 7.2 $\mu_B$,[\onlinecite{boeuf}] the latter two of
which would be more consistent
with a ``divalent Mn'' $d^5_{\uparrow}, d^0_{\downarrow}$ configuration.  
The Curie-Weiss $\theta_{CW} = -160~{\rm K}$ 
suggests large AFM interactions between the Mn moments\cite{endo},
consistent with the AFM ordering.

CuMnSb is metallic, 
but experimentally much less so than the FM compounds in its
class.  Its Drude plasma energy  was reported\cite{kiril} as $\hbar \Omega_p$ = 
1.6 eV compared to 2.6-6~eV for the FM Heusler compounds, 
suggesting a much smaller density of states and/or Fermi velocity, and an 
interband absorption peak was reported at 0.8 eV.\cite{bobo}
For a metal, the
square $\Omega_p^2$ is proportional to N(E$_F$)$v_F^2$ (often interpreted as
an effective carrier density to mass ratio
$(n/m)_{eff}$), and this square is a factor or 3-15 lower 
than its FM counterparts.

The resistivity of CuMnSb
shows a fairly rapid drop below T$_N$, with the samples of Schreiner and 
Brand\~ao\cite{schreiner} dropping from around 170 $\mu \Omega$ cm 
above T$_N$ to 50
$\mu \Omega$ cm at low temperature.  More recent work\cite{thesis} has lowered 
these numbers somewhat: 120 $\mu \Omega$ cm above T$_N$ to 45 
~$\mu \Omega$~cm at low temperature.
Such resistivity drops at magnetic
ordering transitions are common and reflect 
spin scattering that gets frozen out in the ordered phase.  
In contradiction to this inference, and
quite unusual for magnetic compounds, CuMnSb shows little 
magnetoresistance.\cite{thesis,boeuf} The
still large residual resistivity ($\approx~45~\mu\Omega~cm$)\cite{otto}
in the best samples suggests some intersite disorder
or non-stoichiometry (the empty site makes half Heuslers 
susceptible to such defects).  Our work indicates this compound to be
a semimetal, however, so this residual resistivity alternatively may
reflect primarily a low carrier density for a metal.
In both AFM CuMnSb and several FM half Heusler
compounds, it has reported that the resistivity is not T$^2$ at low
temperature,\cite{otto} however Boeuf\cite{thesis} obtains
 essentially a T$^2$ behavior. 

Possibly related to some of these features is the fact that
CuMnSb displays a curious stability of its magnetic order under applied
magnetic fields.  The N\'eel temperature is invariant for fields up to 
14 T.  Whereas many AFMs display metamagnetic transitions in a field,
the induced moment at 5~K and H~=~12 T is only 0.25 $\mu_B$, again reflecting
its imperviousness to applied fields.  In a more recent work Doerr
et al.\cite{doerr} have seen no changes of the antiferromagnetic 
characteristics up to 50~T.  

The remaining evidence about the electronic state of CuMnSb is from heat
capacity (C$_V$) data.\cite{thesis,boeuf}  
The linear specific heat coefficient $\gamma$~=~17~mJ/mol~K$^2$ 
corresponds to a quasiparticle density of states 
N$^*$(E$_F$)~=~7.2~states/eV per formula unit.  
The T$^3$ coefficient of C$_V$ was reported to be
two orders of magnitude larger than the anticipated lattice contribution,
which by itself would suggest it is dominated by soft magnetic fluctuations.
Half-Heusler antimonides have recently attracted attention due to their
large thermopower and other promising thermoelectric properties,\cite{mastro}
but the Seebeck coefficient of CuMnSb has not been reported.
 
Although there has been almost no electronic structure study of CuMnSb,
many other Heusler and half Heusler compounds have been studied rather
thoroughly.  A recent analysis of trends in the electronic structure and 
magnetization of Heusler compounds was been presented by Galanakis,
Dederichs, and Papanikolaou,\cite{galanakis} which references much of the
earlier work.  A recent review concentrating on spintronics applications has
been provided by Palmstrom.\cite{palmstrom} An extensive set of
calculations for many half Heusler compounds has been presented by
Nanda and Dasgupta.\cite{nanda} None of these address Cu-containing
compounds, however.

In this paper we lay the foundations for an understanding of the electronic
properties and magnetism of CuMnSb by reporting local spin density results
for the electronic and magnetic structures, and the equation of state,
for unpolarized, ferromagnetic, and antiferromagnetic alignment of the 
Mn spins.  The results, such as the AFM ordering and the low density of
states at the Fermi level ${\rm N(E_F)}$, are in reasonable
agreement with experimental data, 
so it appears that LSDA is accurate for CuMnSb as it is in most other
intermetallic $3d$ compounds.  The resulting band structure in the AFM
phase is rather unusual, being that of a compensated semimetal 
with normal mass hole bands but heavy mass electron bands.
Possible correlation effects are investigated using the correlated
band theory (LDA+U) method.

\section{Structure and Method of Calculation}
The half Heusler structure is based on the Heusler structural class AT$_2$X
of numerous intermetallic compounds, with space group Fm3m (\#225) whose
point group contains all 48 cubic operations.  
This structure type
can be pictured in terms of an underlying 
bcc arrangement of atomic sites with lattice constant $a$/2, with atom
A at (0,0,0), X at ($\frac{1}{2},\frac{1}{2},\frac{1}{2})a$, and T at
($\frac{1}{4},\frac{1}{4},\frac{1}{4})a$ and 
($\frac{3}{4},\frac{3}{4},\frac{3}{4})a$. Thus the T and X sites lie on the
corner sites of the bcc lattice (alternating), while the A sites comprise the
body center sites.  In the half Heusler structure,
one of the T sites is unoccupied.  The CuMnSb therefore can be viewed as
``MnCu$_2$Sb'' with the Cu site at ($\frac{1}{2},\frac{1}{2},\frac{1}{2})a$
unoccupied.

There are alternative views of the structure.  For example, in AT$_2$X,
the A and X sites form a rocksalt lattice and the T atoms fill the
tetrahedral sites of this lattice.  Alternatively, in half Heusler
ATX, A and T form a zincblende lattice and half of the interstices 
(those neighboring T) are filled with X atoms.  An alternative view of
last choice is the viewpoint of a zincblende lattice formed of T and X atoms,
with A placed at the interstitial site nearest T.  We will study whether
any of this viewpoints is preferable to the others.

The half Heusler structure has space group F${\bar 4}3m$ (\#216) with the
tetrahedral point group.  The crystal structure\cite{forster}
 is fcc, lattice constant $a$ =
6.088~\AA, with Mn (A) at (0,0,0), Cu (T) at
($\frac{1}{4},\frac{1}{4},\frac{1}{4}$) and Sb (X) at 
($\frac{1}{2},\frac{1}{2},\frac{1}{2}$).  
The symmetry of all the sites is ${\bar 4}3m$. 
It will become significant
that in this structure the magnetic atom Mn is coordinated tetrahedrally
by four Cu atoms at $(\sqrt{3}/4)a$~=~2.64~\AA, with the other four 
nearest sites being vacant.  Mn is second-neighbored 
by six Sb atoms at a distance
of $a/2$=3.04~\AA.  The Mn-Cu distance is almost identical to the sum
of their atomic radii (1.35~\AA~and 1.28~\AA, respectively), while the Mn-Sb
distance is 3\% greater than the sum of their radii (1.35~\AA~and 1.59~\AA,
respectively).
The AFM structure consists of alternating (111) planes of Mn atoms
with aligned spins.
The Mn-Mn nearest neighbor distance of 4.31 \AA~assures that direct Mn-Mn
exchange is not a dominant factor in the magnetic ordering.

\begin{figure}[tb]
\psfig{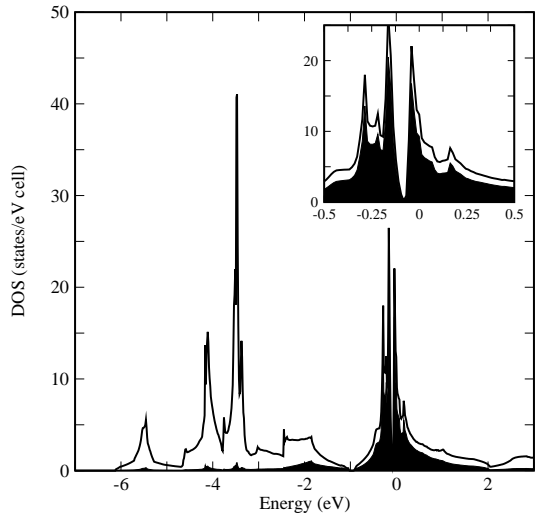}
\caption{Density of states of nonmagnetic CuMnSb, with the Mn $3d$ 
contribution shaded.  The inset shows the anomalous accidental ``zero gap''
that occurs slightly below E$_F$. 
}
\label{PMDOS}
\end{figure}

Two types of band structure methods have been used.  The full-potential 
nonorthogonal local-orbital minimum-basis scheme (FPLO)\cite{fplo1, fplo2}  
was used for scalar relativistic calculations in the 
local density approximation (LDA) for the exchange-correlation energy\cite{lda}.
Cu $3s,3p,4s,4p,3d$, Mn $3s,3p,4s,4p,3d$, and 
Sb  $4s,4p,5s,5p,4d$ states were included as 
valence states. All lower states were treated as core states.
We included the relatively extended semicore $3s,3p$ states of Cu and Mn, 
and $4s,4p$ semicores of Sb
as band states because of the considerable overlap of these 
states on nearest neighbors.
This overlap would be otherwise neglected in the FPLO scheme.
The spatial extension of the 
basis orbitals, controlled by a confining potential $(r/r_{0})^4$, was 
optimized to minimize the total energy.
The self-consistent potentials were 
calculated on a 12$\times$12$\times$12 k mesh in the Brillouin zone,
which corresponds to 116 k points in the irreducible zone.

For the magnetic calculations 
the full-potential linearized augmented-plane-wave (FLAPW) method
as implemented in the Wien2k code\cite{wien2k} was used.
The $s$, $p$, and $d$ states were treated using
the APW+lo scheme \cite{sjo00}, while the standard LAPW
expansion was used for higher $l$'s.  
The basis size was determined by $R_{mt}K_{max}=7.0$.
The space group of the AFM structure is R$3m$ (\#160).  The results
we present result from use of the Perdew-Burke-Ernzerhof generalized
gradient approximation form\cite{pbe96} of exchange-correlation 
functional, but for the bands and density of states the results are
similar to those from the local density approximation.

\section{Results of Electronic Structure Calculations}
\subsection{Paramagnetic phase}  

The (unstable) paramagnetic electronic structure 
of CuMnSb is unusual.  
The density of states (DOS), shown in Fig. \ref{PMDOS}, exhibits
a narrow peak around the Fermi level E$_F$, with an extremely narrow
valley slightly below E$_F$.  The five $3d$ bands that in metallic Mn
are spread over $\sim$5 eV are here confined to a 1 eV region 
centered on E$_F$.
The split peak is the result of a gap between the disjoint
``valence'' bands and ``conduction'' bands, but the gap is indirect and
accidentally zero.  CuMnSb, with an odd number of valence electrons in the
primitive cell, cannot be a nonmagnetic insulator, and the occupied states
include one electron in what we are referring to as the conduction bands
(those above the tiny gap).  The upper valence and lower conduction bands 
show a {\it direct} gap of $\sim$0.1 eV around the zone edge L, W, and U
points.  On either side of the gap the bands are flat over most of the
surface of the Brillouin zone, and analysis shows that 
Mn $t_{2g}$ character dominates the 
bands on both sides of the gap (especially above). 

The Heusler and
half Heusler structures tend to give rise to such gaps, which is the origin of
the various occurrences of half metallic ferromagnetism in this
structure.  One should take the point of view that CuMnSb contains only
one transition metal atom, since the Cu $3d^{10}$ shell is completely
filled and thus inert.
Therefore the Cu $3d$ states are not involved 
in a double-peaked complex of narrow bands at E$_F$.  
Ironically, we find that the 
presence of Cu in the
compound is the key to the peaked Mn $3d$ DOS and possibly to
the peculiarity of the magnetism.
The Mn atom sits at the center of a minicube in which it is coordinated
tetrahedrally by four Cu atoms, and also tetrahedrally by four unoccupied
sites in the half Heusler lattice.  Hence the Mn $3d$ states have only the $p$
states of the second neighbor Sb atoms (along the cubic axes) and some
weak Cu $4s$ character to 
hybridize with and hence to broaden.  This coupling is weak, as reflected
in the narrow Mn DOS.  To establish the effect of Cu quantitatively for
states around the Fermi level, we also carried out a calculation of 
$\bigcirc$$^{+1}$MnSb, which denotes CuMnSb with the Cu atom 
removed but leaving its
$4s$ electron (which is simply added to the system).  The change is very 
small near E$_F$, the one effect being that the threefold level 
above E$_F$ at the zone center
is shifted upward by 0.6 eV by the presence of Cu. 

\begin{figure}[tb]
\noindent
\psfig{figure=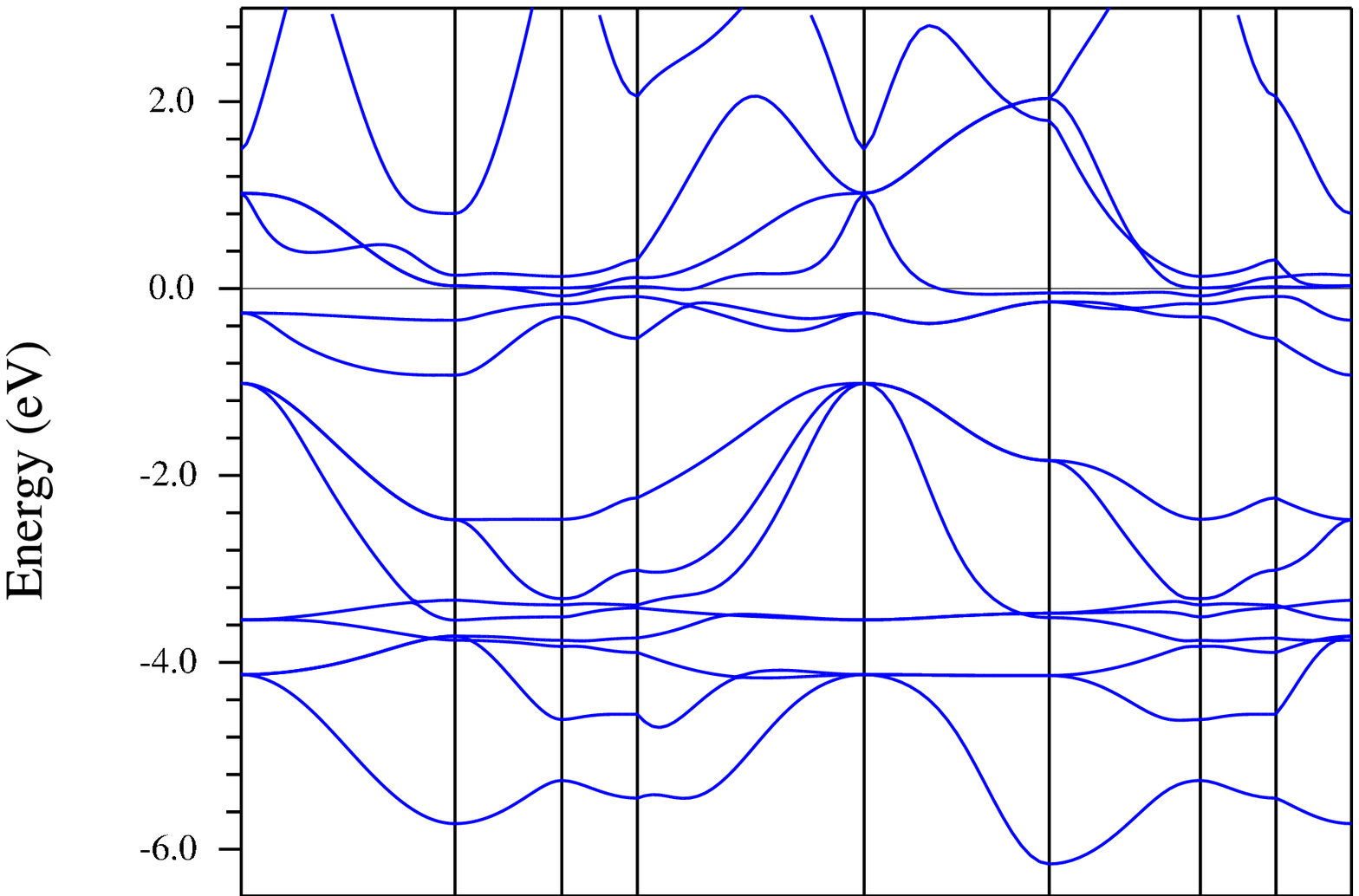,width=9.0cm}
\noindent
\psfig{figure=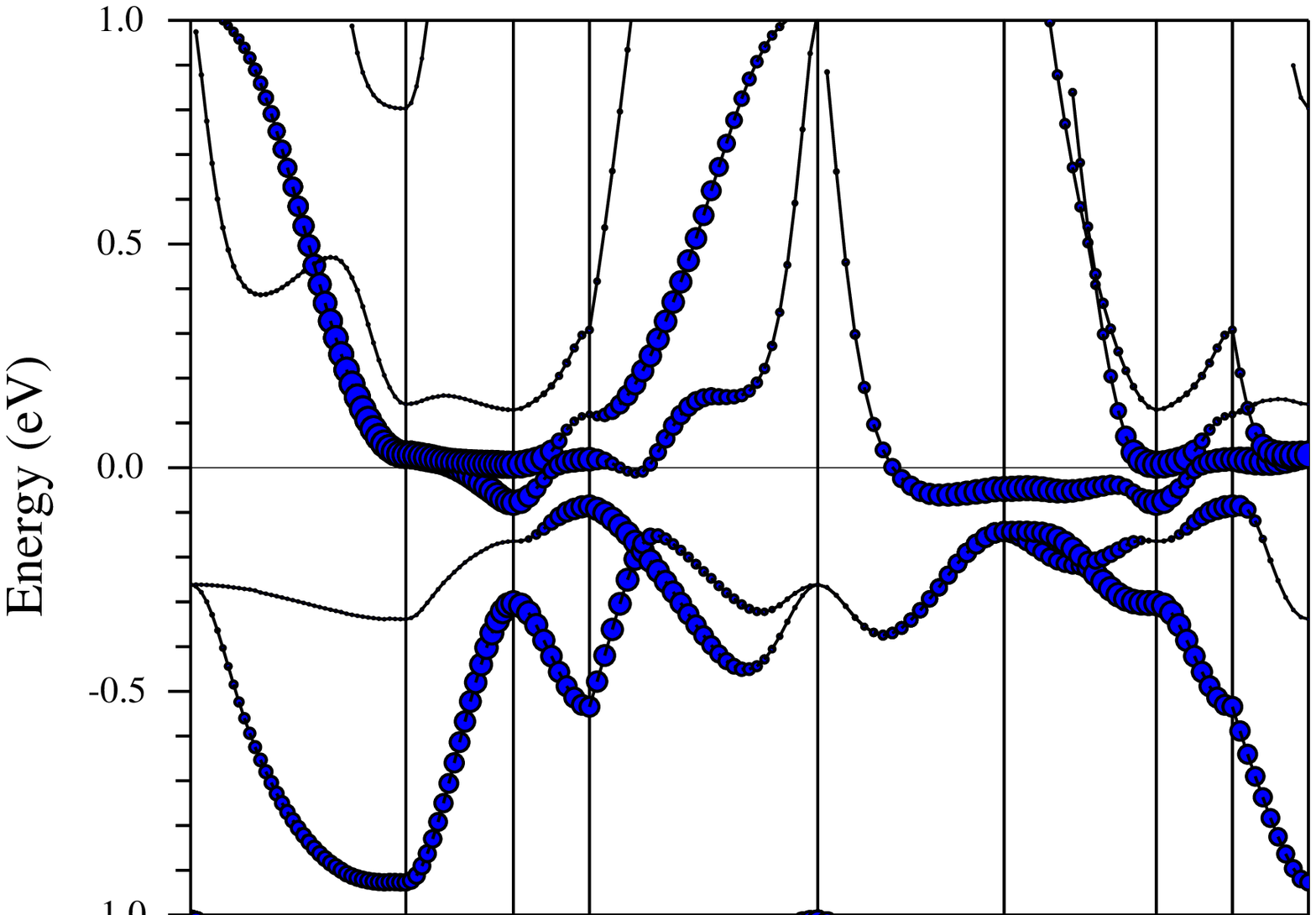,width=9.0cm}
\caption{Top panel: nonmagnetic bands of CuMnSb.  
Notice that the 
two bands in the -1 eV to E$_F$ region are disjoint from not only the lower
bands, but also the near-lying bands above that have the same Mn $3d$ 
character.  The indirect gap from K to W is accidentally almost 
zero.  Bottom panel: 
blowup of the bands very near the Fermi level
emphasizing the bands with Mn $t_{2g}$ character with larger symbols. 
Of these five bands, the ones appearing here as thin lines have $e_g$
character.
}
\label{supercell}
\end{figure}

Whether or not the paramagnetic phase has a Stoner instability, it is clear
from the huge DOS peaks that a FM state with finite moment is
likely to be energetically favorable to the paramagnetic state. 
The calculated Fermi level DOS is N(E$_F$)= 11.68 states/eV per formula unit,
which corresponds to a linear specific heat coefficient (without any
many body enhancement) of 27.5 mJ/mol-K$^2$.
Assuming the usual value of
Stoner interaction parameter $I_{Mn}~\approx~0.75$~eV leads to 
N(E$_F$)$I$~$\sim$~9, reflecting a very strong Stoner instability (which
occurs anytime N(E$_F$)$I$ $\geq$ 1).

\subsection{Ferromagnetic alignment of Mn spins} 
\begin{figure}[tb]
\psfig{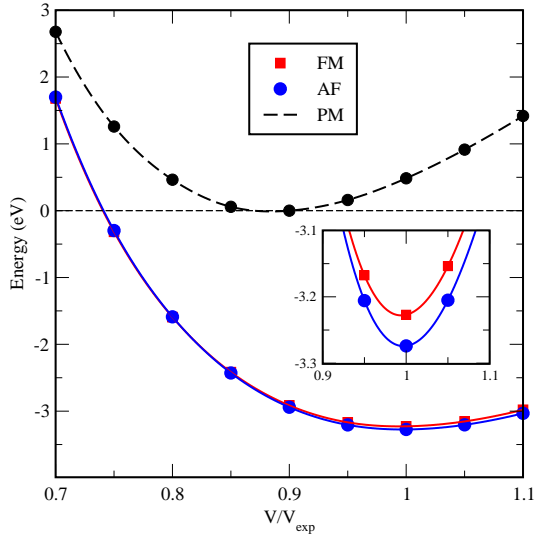}
\caption{Calculated equation of state for nonmagnetic (upper curve) and
magnetic (lower curves) CuMnSb.  On this scale the FM and AFM differences 
cannot be distinguished.  Note the extremely large magnitude of the 
energy gain from magnetism, and the very large ``magnetic expansion''
corresponding to almost 4\% in the lattice constant.
}
\label{EOS}
\end{figure}

Calculation shows that either FM or AFM order lowers the 
energy by more than 3 eV/Mn compared to a nonmagnetic configuration,
reflecting the energy gain simply from local moment formation.  
In addition, there
is an energy difference arising from the type of spin alignment.
In Fig. \ref{EOS} the equation of state is
pictured for the paramagnetic, FM, and AFM phases.  
The energy difference between FM and AFM phases is about
50 meV, a factor of 60 smaller than the energy gain from moment formation.  
Hence the Mn moment will be very robust in this compound, independent of
degree or type of order.
Also noteworthy is the large difference in 
volumes, the paramagnetic
equilibrium being about 12\% smaller in volume (4\% smaller lattice constant)
than the magnetic phases.

\begin{figure}[tb]
\psfig{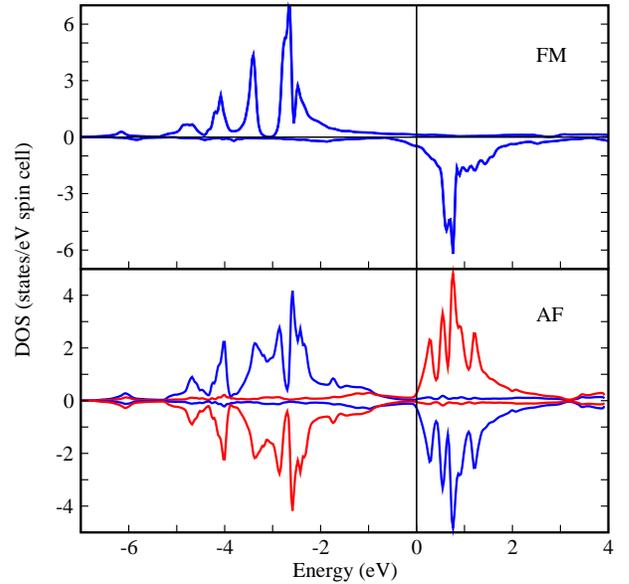}
\caption{Comparison of the Mn partial densities of states for the
magnetic (FM and AFM ) configurations.  For FM alignment there is some
occupation of of the minority bands that is evident in the figure.  
Though less obvious, there is also minority occupation in the AFM, and
the moment is 4 $\mu_B$.
}
\label{3DOS}
\end{figure}

Before moving on to discuss the more stable AFM electronic structure, we
analyze briefly the FM results.  The exchange splitting of the Mn $3d$ 
states is 3.5 eV, thus the FM bands look nothing like rigidly shifted 
replicas of the paramagnetic bands, as is evident from the DOS 
pictured in Fig. \ref{3DOS}.  For the unpolarized bands up and down bands
are identical and both are about half-filled; after spin polarization, the
majority are filled and the minority are nearly empty.
The large exchange splitting is characteristic
of the large moment, 3.9 $\mu_B$ in the Mn sphere, 0.1 $\mu_B$ from the 
Cu sphere, and 0.4 $\mu_B$ in the interstitial, 
for a total of
4.4 $\mu_B$ per formula unit.  The interstitial contribution is perhaps 
mostly from Mn tails, since the Mn majority states are clearly entirely
filled (Fig. \ref{3DOS}) but there is some small filling of the Mn
minority $3d$ states that will reduce the moment somewhat, {\it i.e.}
5 ($\uparrow$) -0.6 ($\downarrow$) = 4.4 $\mu_B$.  

Once the moment is formed, the majority 
Mn $3d$ bands overlap the Cu
$3d$ bands in energy and mix strongly, and together form two nearly 
dispersionless complexes of 
bands along {$\Gamma$-A} at -3.4 eV and -2.75 eV.  
This repulsive hybridization with the
closed $3d$ shell of Cu should contribute to the large lattice expansion
noted above, which is larger than that due to the conventional volume
expansion due to moment formation.  The minority Mn $3d$ states 
hybridize only with Sb $5p$ or Cu $4s$, and form a narrow set of bands 0.8
eV above E$_F$ that is only 0.4 eV wide (with some tailing).  
The resulting Fermi level DOS is quite low, 
N(E$_F$)~=~0.505~(up) + 
0.845~(down) = 1.35 states/eV per formula unit.
While this FM alignment is metastable, 
the EOS calculations show that AFM alignment
is more stable, in agreement with experiment.
Between the exchange split Mn $3d$ states lie bands with 2-3 eV 
dispersion that have primarily Sb $5p$ character.

\subsection{Antiferromagnetic alignment of Mn spins}
We obtain an AFM state for CuMnSb whose Mn $3d$ density of states is 
shown in Fig. 
\ref{3DOS}.  The magnetic moment lies almost entirely on the Mn atom,
which is essentially a fully occupied majority and some minority occupation,
with moment $m$ = 3.9 $\mu_B$ 
within the Mn sphere of radius 2.2 bohr, in excellent agreement with
the measured ordered moment.\cite{helmholdt,forster}  The
Mn exchange splitting is 3.5 eV, the same as for the FM case and again
consistent with the large moment.

\begin{figure}[tb]
\psfig{figure=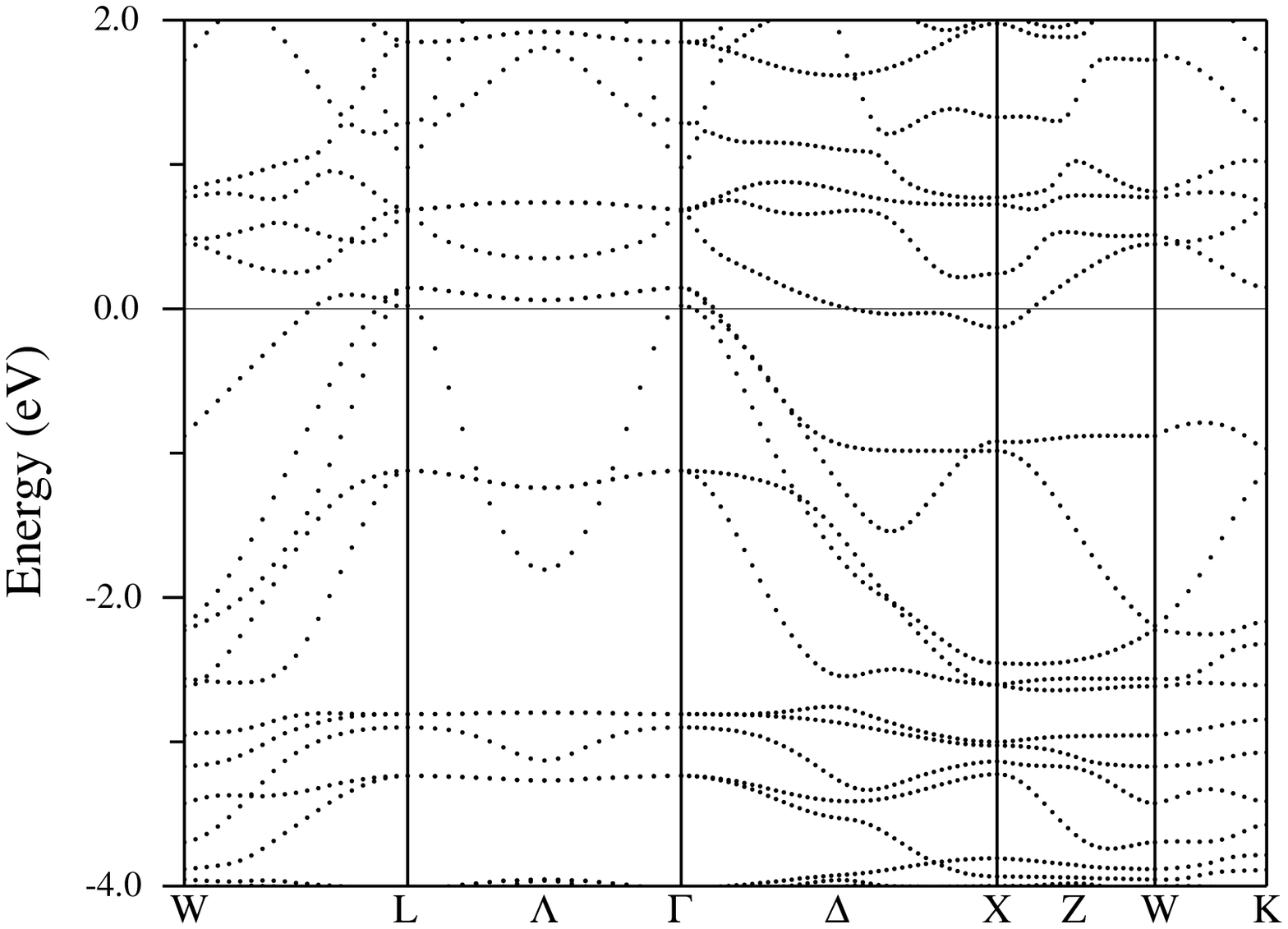,width=9.5cm}
\psfig{figure=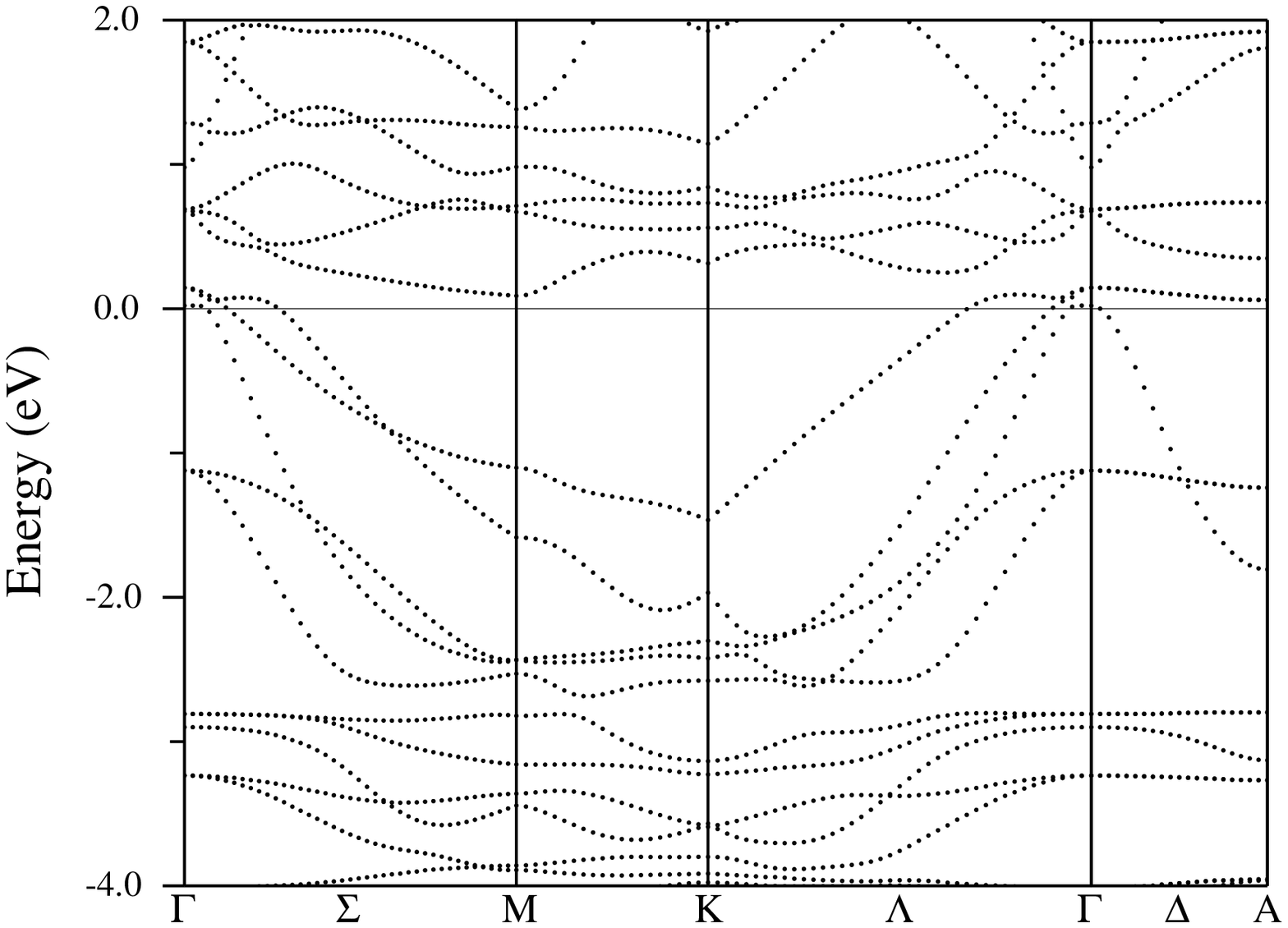,width=9.5cm}
\caption{Bands of antiferromagnetic CuMnSb, plotted along lines in both the
underlying fcc symmetry (top) and in the
hexagonal AFM symmetry(bottom).
The $\Gamma$-L direction shown is along the hexagonal
axis of the AFM cell.  As for the paramagnetic phase, the bands (primarily)
below E$_F$=0 (valence) are disjoint from those above (conduction), but
each overlap E$_F$ by 0.1 eV.  The exchange split Mn $3d$ bands lie 
around -3 eV and +0.5 eV.
}
\label{AFMbands}
\end{figure}

Fig. \ref{3DOS} reflects also a very low density of states at E$_F$,
whose value is N(E$_F$) = 1.38 states/eV per formula unit.
As in the paramagnetic phase, the valence and conduction
bands are disjoint, as can be seen in Fig. \ref{AFMbands}.  There are two 
small, and one larger, Fermi surface hole cylinders along the 
$\Gamma$-A $(0,0,k_z)$ 
line in this figure, 
which corresponds to one
$\Gamma$-L $<111>$ direction in the original fcc zone.  These
holes are compensated by electron pockets at the X point of the fcc
zone. 

\subsection{Possible correlation corrections: LDA+U}
It is  common for the local spin density approximation to
work very well for transition metal intermetallics and in compounds with
pnictides, and it has been used extensively in the Heusler and half-Heusler
classes.  Given the relative chemical isolation of the Mn moment in
CuMnSb, however, one may wonder if there are residual correlations effects that
should be accounted for, such as with the LDA+U method.\cite{ldau}
LDA+U is known
to do a good job of modeling the band equivalent of the Mott insulator in
several systems, mostly oxides.  CuMnSb is a metal, and in fact the 
calculated LSDA DOS seems to be at least qualitatively
reasonable compared to the quasiparticle
DOS obtained from the linear specific heat coefficient.  Still, it is
important to know if correlation corrections could change the results
appreciably.

\begin{figure}[tb]
\psfig{figure=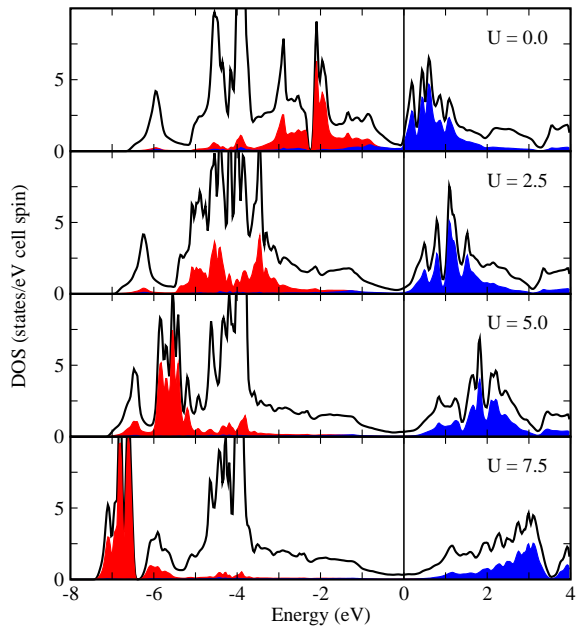,width=9.5cm}
\caption{(color online) Total and Mn $3d$-projected (filled-in regions)
density of states of CuMnSb, using
the LDA+U method with U = 0 (LDA), 2.5, 5.0, and 
7.5 eV (top to bottom).  The only 
noteworthy change is the lowering of occupied, and raising of unoccupied,
Mn $3d$ states.  The small DOS around the Fermi level remains.
}
\label{LDAUDOS}
\end{figure}

We have applied the LDA+U functional as adapted to the linearized
augmented plane wave method by Shick {\it et al.}\cite{shick} 
and implemented in the
Wien2k code to investigate this question.  The LDA+U method requires the
values of the Mn $3d$ Coulomb repulsion $U$ and exchange $J$ constants.
Because we use the ``fully-localized limit'' form\cite{fll} of functional, we
make the common replacement $U \rightarrow U_{eff}
\equiv U-J$, and have tried the values $U$ = 2.5, 5.0, and 7.5 eV.  
The resulting total
and Mn-projected DOS are shown in Fig. \ref{LDAUDOS}. 
The occupied and unoccupied Mn 3$d$ states behave as expected, each
moving away from the Fermi level as U is increased.  Except for reduction
of hybridization due to this shift
of $3d$ states, the bands are changed little. In particular, the 
band overlap describing a compensated semimetal remains, however it is increased
by 0.2 eV.  The calculated
values of N(E$_F$) for U = 0, 2.5, 5.0, and 7.5 eV are 1.38, 0.76, 0.66,
and 0.66 states/eV cell (both spins), respectively.  Thus  the effect
of including $U$ is to {\it reduce} the {\it band} mass by as much as 50\%,
due to the reduction of the minority Mn $3d$ character in the conduction
bands.  The corresponding values of the Mn moment (inside the muffin-tin
sphere) are 3.87, 4.20, 4.40, 4.52 $\mu_B$ respectively.

\begin{figure}[tb]
\psfig{figure=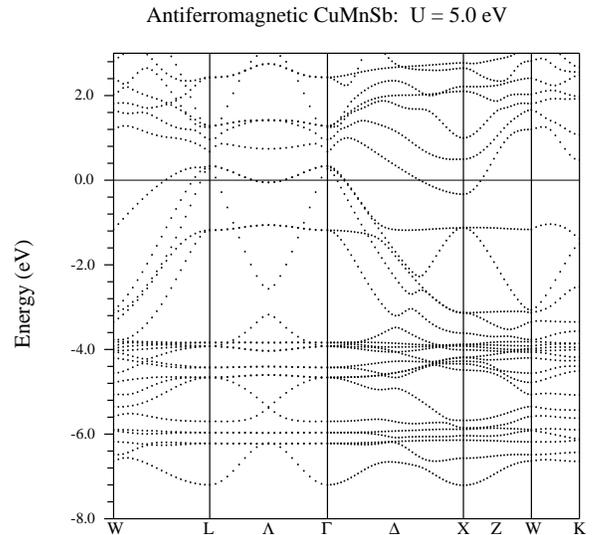,width=9.5cm}
\caption{Bands of CuMnSb using the LDA+U method, with U = 5 eV, 
The bands are plotted along the underlying fcc lines (compare to the
upper panel of Fig. 5)
so both electron and hole pockets are evident.  The most noticable 
effect of including $U$ is the lifting of minority $3d$ bands to the
ragne 1-3 eV, and almost completely eliminated $3d$ character at
the Fermi level.
}
\label{LDAUbands}
\end{figure}

The band structure for U = 5 eV (which is probably an upper limit for the
value of U for Mn in an intermetallic compound) is shown in 
Fig. \ref{LDAUbands}.  The increased band overlap, relative to U=0, 
results in larger Fermi surfaces for both electrons
and holes.  There is a single electron ellipsoid (at the point that
would be X in the cubic Brillouin zone), 
while there are three closed hole surfaces around both the $\Gamma$ 
and A points of the hexagonal Brillouin zone.

The separation (disjointness)
between the conduction and valence bands becomes clearer
when $U > 0$ is included (Fig. \ref{LDAUbands}), which reflects some
sort of `bonding - antibonding' separation that is not very clear in
terms of interaction of atomic orbitals.   
One might then be concerned about the `band gap problem' in the local
density approximation, for which a gap separating bands of distinctly
different character is usually underestimated.  One approximate remedy
for this problem is the GW approximation, in which a nonlocal dynamically
screened exchange interaction is included.\cite{gw}  The projected 
density of states however shows
no clear distinction between the characters of the $s-p$ bands
below and above the gap, so the self-energy correction is not likely to
be large.  Since the band structure is that of a heavily doped
self-compensated semiconductor (whether with $U$ = 0 or $U$ = 5 eV),
any correction will have some effect on the carrier density and on
the value of N(E$_F$).

\section{Discussion and Summary}
In this local-density based study of the electronic and magnetic structure of
the half-Heusler magnet CuMnSb we have found AFM alignment to be 
energetically favored, consistent with experiment.  The moments, for which
the LSDA value of 4 $\mu_B$ is in excellent agreement with the 
observed ordered moment, are robust
independent of relative orientation.
The characters of occupied states indicate that a Cu$^{1+}$Mn$^{2+}$Sb$^{3-}$
configuration may be useful starting point
for characterizing the electronic structure, although within LSDA
there is clearly
some minority $3d$ occupation.
The electronic structures of all phases are representative of narrow Mn $3d$ 
bands in the midst of much broader bands consisting of Cu $4s$, Sb $5s5p$,
and, further above, Mn $4s$, and are consistent with earlier calculation of
the density of states.\cite{poland}

The (LSDA) energy difference of 50 meV/Mn between AFM and FM alignment of spins
translates, in a nearest neighbor exchange coupling model, to $J~S^2$=8.3 meV
which with $S=\frac{5}{2}$ gives $J$= 1.3 meV = 15 K.
Such a picture of coupling is probably not of quantitative
value for CuMnSb, because here interactions between Mn
spins will be mediated by a low density of semimetal carriers, leading
to coupling with several neighboring shells.
There will be two distinct interactions,
one through the heavy conduction electrons, another through the light
valence band holes.  This situation has much in common with EuB$_6$, which
a recent study\cite{kunes} has shown to have two such local-moment to
itinerant-state interactions, with a different sign of the
on-site Kondo coupling for valence and conduction bands.

The calculated LDA value of N($E_F$) = 1.38 states/eV per formula unit
 can be compared with
the dressed one obtained from the linear specific heat coefficient
N$^*$(E$_F$)= 7.2 states/eV per formula unit to obtain a dynamic thermal mass enhancement
$\lambda$ = $m^*/m$ - 1 $\approx$ 4.2. This result is indicative of a large
mass enhancement due to spin fluctuations.  For $U$ = 5 eV, N(E$_F$) 
drops to 0.66 states/eV per formula unit and thermal mass enhancement 
increases to $\lambda$ $\approx$ 10.

This nonmagnetic electronic structure is reminiscent of paramagnetic FeSi, 
in which there is a tiny gap\cite{mattheis} 
of 0.13 eV between flat bands that are of
mostly Fe character.
In FeSi the Fe moment is very obvious in the Curie-Weiss susceptibility
at higher temperature but becomes compensated at low temperature and the
ground state is paramagnetic.  In CuMnSb the electron count is such that
there is one electron above a similar looking gap.  The other clear 
difference is that the Mn atom remains strongly 
polarized in CuMnSb, while the Fe atom in FeSi loses its magnetism
at low temperature.

Our study of the electronic structure of CuMnSb reveals that the Mn $3d$
states are virtually chemically isolated from the environment, leading
to very narrow $3d$ bands and a strong Stoner instability.  In the
(observed and predicted) antiferromagnetic state, the system 
is semimetallic, with 
rather normal mass Sb $5p$ hole bands.  These bands overlap a single 
electron band, whose character depends on whether LDA is accurate or if
correlations are appreciable.  Within LDA, the electron band contains 
some minority Mn $3d$ character, while this mixing rapidly disappears if
U is appreciable.  Comparing the
calculated N(E$_F$) to the specific heat $\gamma$ indicates a 
mass enhancement $m^*/m \sim 5$ (LDA) or as much as $m^*/m \sim$ 11
(U$\geq$ 5 eV).    The observed temperature dependence of the transport
coefficients show no evidence of heavy fermion behavior, however.

In the transition metal based Heusler and half-Heusler compounds ferromagnetism
is common; CuMnSb is practically the only antiferromagnet.  Our calculations
indeed predict that AFM has the lower energy, although exploring the spin
coupling mechanisms has not been the purpose of this work.  The magnetic
interactions presumably are mediated separately through the
hole and electron bands, indeed it could be the case that the two mechanisms
compete, as has been uncovered recently\cite{kunes} for EuB$_6$.

The role of the Cu atom is as a spacer and donor of an electron.  Cu is
nevertheless crucial, since without it MnSb crystallizes in the hexagonal NiAs
structure and is ferromagnetic rather than antiferromagnetic.  Electronic
structure studies of the NiAs-structure phase as well as the possible
zincblende phase have been reported previously.\cite{continenza} 

\section{Acknowledgments}
We gratefully acknowledge illuminating discussions with C. Pfleiderer, who
also provided a copy of Ref. \onlinecite{thesis}.
This work was supported by DOE grant DE-FG02-04ER46111.
R.W. is member of {CONICET}
(Consejo Nacional de Investigaciones Cient\'{\i}ficas y T\'ecnicas, Argentina).


\end{document}